\documentclass{svjour3}                     

\smartqed  

\usepackage{graphicx}

\begin{document}

\title{Instruments of RT-2 Experiment onboard CORONAS-PHOTON 
and their test and evaluation I: Ground calibration of RT-2/S and RT-2/G
\thanks{This work was made possible in part from a grant from Indian Space Research Organization 
(ISRO). The whole-hearted support from G. Madhavan Nair, Ex-Chairman, ISRO, who initiated the 
RT-2 project, is gratefully acknowledged.}}

\titlerunning{RT-2 payloads onboard CORONAS-PHOTON I: RT-2/S and RT-2/G}

\author{Dipak Debnath \and Anuj Nandi \and A. R. Rao \and J. P. Malkar \and M. K. Hingar \and 
T. B. Kotoch \and S. Sreekumar \and  V. P. Madhav \and Sandip K. Chakrabarti}

\authorrunning{Dipak Debnath et. al. } 

\institute{Dipak Debnath, Anuj Nandi$^+$, T. B. Kotoch \at
Indian Centre for Space Physics, 43 Chalantika, Garia Station Rd., Kolkata 700084, India\\
              \email{dipak@csp.res.in; anuj@csp.res.in; tilak@csp.res.in}\\
($+$: Posted at ICSP by Space Science Division, ISRO Head Quarters, Bangalore, India)
\and
A. R. Rao, J. P. Malkar, M. K. Hingar, V. P. Madhav \at
Tata Institute of Fundamental Research, Homi Bhabha Road, Colaba, Mumbai 400005, India\\
              \email{arrao@tifr.res.in; jpm@tifr.res.in; mkhingar@tifr.res.in; 
vaibhav1881@gmail.com}\\
\and
S. Sreekumar \at
Vikram Sarabhai Space Centre, VRC, Thiruvananthapuram 695 022, India.\\
              \email{sreekumar\_s@vssc.gov.in}\\
\and
Sandip K. Chakrabarti \at
              S.N. Bose National Centre for Basic Sciences, JD Block, Salt Lake, Kolkata 700097, 
India \\
(Also at Indian Centre for Space Physics, 43 Chalantika, Garia Station Rd., Kolkata 700084)\\
              \email{chakraba@bose.res.in}            \\
}

\date{Received: date / Accepted: date}
\maketitle

\begin{abstract}

Phoswich detectors (RT-2/S \& RT-2/G) are major scientific payloads of the RT-2 Experiment 
onboard the CORONAS-PHOTON mission, which was launched into a polar 
Low Earth Orbit of around 550 km on 2009 January 30. These RT-2 instruments are designed and 
developed to observe solar flares in hard X-rays and to understand the energy transport processes 
associated with these flares. Apart from this, these instruments are capable of observing Gamma
Ray Bursts (GRBs) and Cosmic diffuse X-ray background (CDXRB).
Both detectors consist of identical NaI(Tl) and CsI(Na) scintillation crystals 
in a Phoswich combination, having the same diameter ($116$ mm) but different thicknesses. The 
normal working energy range is from 15 keV to 150 keV, but may be extendable up to $\sim$ 1 MeV. 
In this paper, we present the RT-2/S and RT-2/G instruments and discuss their testing 
and calibration results. We used different radio-active sources to calibrate
both detectors. The radio-active source $^{57}$Co (122 keV) is used for onboard calibration 
of both instruments. 
During its lifetime ($\sim$ 3-5 years), RT-2 is expected to cover the peak of the $24^{th}$ solar cycle.


\end{abstract}

\section{Introduction}

The RT-2 Experiment onboard the CORONAS-PHOTON mission (Kotov et al. 2008, Nandi et al. 2009) is 
a low energy $\gamma$-ray telescope. 
The experiment consists of three scientific payloads (two Phoswich detectors and 
one solid-state imaging device) and one processing electronic payload. The Phoswich 
detectors are RT-2/S and RT-2/G, which will work in the energy range from 15 to 150 keV, 
extendable upto 1 MeV. The solid-state imaging payload RT-2/CZT, consists of 
three CZT (Cadmium Zinc Telluride) detector modules and one CMOS (Complimentary Metal 
Oxide Semiconductor) detector (Kotoch et al. 2010, Nandi et al. 2010). 
Data from all these three detectors are processed by an electronic device, namely, RT-2/E 
(Sreekumar et al. 2010), which also interfaces with the Satellite system.

All three payloads are placed outside the hermetically sealed vessel of the satellite 
and co-aligned to the Sun pointing axis. The instruments have different field of view 
(FOV) of $4^\circ \times 4^\circ$ for RT-2/S, $6^\circ \times 6^\circ$ for RT-2/G 
and varying FOVs, from $6' \times 6'$ to $6^\circ \times 6^\circ$, for RT-2/CZT 
(see Nandi et al. 2010). Though the experiment is dedicated to solar studies of high energy 
phenomena, this configuration also allows the study of a large number of galactic and 
extra-galactic X-ray sources (within $6^\circ$ around the ecliptic plane below 100 keV and half 
the sky in the 100 keV to 1 MeV range -- the latter being due to the fact that the collimators 
themselves become transparent in this range). Hard X-ray detectors on board the satellite with 
different FOV will also help in the measurement of the spectrum of cosmic X-ray background 
(Sarkar et al. 2010).

The paper is divided into the following Sections. In \S 2, we discuss the main scientific goals 
of the RT-2 payloads. In \S 3, we describe the RT-2/S and RT-2/G instruments characteristics 
as well as the signal processing methods used. In \S 4, we present the testing set-up 
and the results on the test and evaluation of these instruments at the laboratory. Finally, in 
\S 5, we present our concluding remarks.
 
\section{Goals and Scientific objectives of RT-2/S \& RT-2/G}

Being the nearest star, the Sun occupies a special place in astrophysical studies as it can be 
more closely observed than any other star. Several dedicated satellites (RHESSI, SOHO, GOES etc.) 
(Lin et al. 1998, Domingo et al. 1994, Harvey 2007)
have been launched to understand its behavior in a wide energy band of electromagnetic radiation. 
Even though its surface temperature is only $\sim$6000 K, it emits X-rays and $\gamma$-rays up to 
a few MeV. This is because of rapid magnetic reconnection which produces energetic solar 
activities. Apart from the thermal electrons which obey the Maxwell-Boltzmann distribution, the 
charged particles, especially electrons, are accelerated by shocks and acquire a non-thermal 
(power-law) distribution. These non-thermal electrons emit energetic synchrotron emission. 
The Russian satellite CORONAS-PHOTON which was launched on January 30, 2009
is the last in the CORONAS series of satellites dedicated to study these types of solar activities.
The RT-2 Experiment along with the high energy gamma-ray instrument (NATALYA-2M; Kotov et al. 2008)
and other instruments onboard CORONAS-PHOTON (Kotov et al. 2004, 2008) will cover a wide band of 
energy from UV to $\sim$ 2000 MeV. One of the main goals of this mission is to understand the 
energy transport processes on the solar surface. It is may be due to plasma oscillations, or 
pinching or sausage instabilities in the magnetic field. Apart from this, the main objective 
of the RT-2 Experiment is to study (i) the time resolved hard X-ray spectra of solar flares, 
(ii) the galactic and extra-galactic sources near the ecliptic plane, (iii) the gamma ray 
bursts (GRBs) and (iv) diffuse cosmic X-ray background, in a wide energy range. 

Since the mission is expected to be in orbit for the next 3-5 years, it is expected to cover 
the $24^\mathrm{th}$ solar cycle with the peak expected in 2012. During a peak, the number of 
sunspots and flares go up drastically. The magnetic activities associated with the sunspots 
which dramatically influence the UV/soft X-ray emissions are therefore expected to peak very soon.
The accompanying soft X-rays also have an impact on the Earth upper atmosphere. 
As it is known extreme ultraviolet radiation and the soft X-ray photons basically 
form the different layers of the ionosphere (upper atmosphere), 
which consists of C, D, E and F layers based on their ion-electron densities.                             
During a solar flare, the electron number density and the 
reflection height change on the sunlit side of the Earth and Very Low Frequency (VLF) radio waves
reflected by the upper atmosphere show a sharp rise in intensity.
From the {\it ab initio} calculation, one can find out the source
strength (solar flare) using the ion-electron density distribution in the D-layer.


Solar flares are the most powerful emission in the entire solar system. They could release 
energies from $10^{32}$ to $10^{33}$ ergs over a timescale extending from $100$ to $1000$ seconds.
It is also found that during a flare, the electrons are accelerated to $10 - 100$ keV using a 
significant fraction of this energy budget. From RHESSI observation, it has been established 
that the particle acceleration and the energy release process (Lin et al. 2003) are linked 
together. The instruments RT-2/S and RT-2/G offer a unique opportunity to investigate the
energetics of solar flares, and hence the particle acceleration, in the $15-150$ keV energy 
range in `phoswich' mode (i.e., spectral data from the NaI crystal) and in the $\sim 100$~keV -- $1$~MeV 
energy range in the spectroscopic mode (i.e., spectral data from CsI crystal using both G1 and G2
amplifiers) during the next solar cycle. Apart from these two `scientific' modes, RT-2/S and RT-2/G
would be operated in two other modes, namely, the Normal mode and the Event mode. Details of these 
two operational modes are discussed in Section 3.4.

At the same time as the RT-2/S and RT-2/G observations will be carried about, the RT-2/CZT 
instrument will enable imaging the hard X-ray solar flares with high spectral and spatial 
resolution in the 20--150 keV energy range (Kotoch et al. 2010). 

Other than this, RT-2 will also focus on the large-scale distribution of magnetic flux on the 
Sun. The magnetic energy stored at the solar surface may be released gradually or explosively 
depending on the dynamics of the magnetic field. This released energy heats up the 
``quiet'' corona and also powers the solar wind. Search for nano-flares on the solar surface could 
be of interesting prospect, as both RT-2/S and RT-2/G instruments can operate in high resolution 
timing mode of 0.1 sec and on rare occasions (depending on the memory availability) in the 
time-tagged event mode with a time resolution of 0.3 milliseconds. 
The transient bursts on the solar surface for short durations with significant energy budget 
($\sim 10^{24}$ ergs) are termed as `nano-flares' (Parker 1998), as compared to the largest solar 
flares of energy $\sim 10^{33}$ ergs. It has been observed from several satellite data that
the time-scale variabilities of nano-flares are of $\sim 1.5 - 52$ sec 
(Hudson 1991, Shimizu \& Tsuneta 1997).

It is important to study `nano-flares' for several reasons. For several decades, it was a 
mystery, why the temperature of the solar corona ($>~10^6$ K), the Sun's outer atmosphere, is much 
higher than the temperatures near the Sun's surface (e.g. chromosphere temperature
$\sim 2\times 10^4$ K). But observations made by Japan's Solar-A (Yohkoh) and Solar-B (Hinode) 
satellites revealed that nano-flares play an important role in the coronal (active-region) 
heating process (Hudson 1991, Shimizu \& Tsuneta 1997, Reale, McTiernan \& Testa 2009). Another 
important aspect is to study the distribution law of energy dependence occurrence rate of
nano-flares. 



The RT-2 measurements with other mission payloads' data sets will be useful to infer 
the composition, flux and anisotropy of energetic ions and electrons with high spatial and 
temporal resolution.

\section{Description of the RT-2/S \& RT-2/G instruments}

\subsection {Phoswich Detector}

RT-2/S and RT-2/G payloads consist of NaI(Tl) and CsI(Na) crystals mounted in a
scintillator-phoswich assembly viewed by a photomultiplier tube (PMT). This entire 
assembly is procured from M/S Scionix Holland BV, The Netherlands. The 
NaI crystal is $3$~mm thick and $116$~mm in diameter. The CsI crystal is tapered 
to provide mechanical stability and at the top side it has a diameter of $116$~mm. 
The thickness of the CsI crystal is $25$~mm. The two crystals are optically coupled 
and hermetically sealed with an entrance window on the NaI(Tl) side and a viewing 
glass window coupled to the PMT through an appropriate light guide on the CsI(Na) side. 
The Phoswich combination of NaI and CsI crystals has good hard X-ray absorption properties 
due to the relatively high effective atomic number (32 for NaI and 54 for CsI crystal) and 
efficient optical light emission (415 nm wavelength emission from NaI and 420 nm wavelength 
emission from CsI) (Knoll 1999). The NaI(Tl) crystal is sensitive to X-ray photons of 
$15 - 150$~keV, whereas CsI(Na) is effective for high energy X-ray/$\gamma$-rays of 
$30 - 1000$~keV as well as to charged particle background. 

Another important use of both crystals in the Phoswich mode is for the background rejection. 
The light signal from the CsI(Na) crystal has a different scintillation decay time (650 ns) 
than that from the NaI(Tl) crystal (250 ns) and this distinction is used to eliminate 
non X-ray background in the NaI crystal. The charge particle background will be removed by 
the upper energy threshold. The Phoswich assembly is viewed by a $76.2$~mm diameter 
photomultiplier through a $10$~mm thick light guide. A schematic drawing of `phoswich' 
assembly of NaI(Tl) and CsI(Na) crystals with the PMT is shown in Figure 1.

\begin {figure}[h]
\vbox{
\vskip 0.0 cm
\centering{
\includegraphics[width=4.5cm]{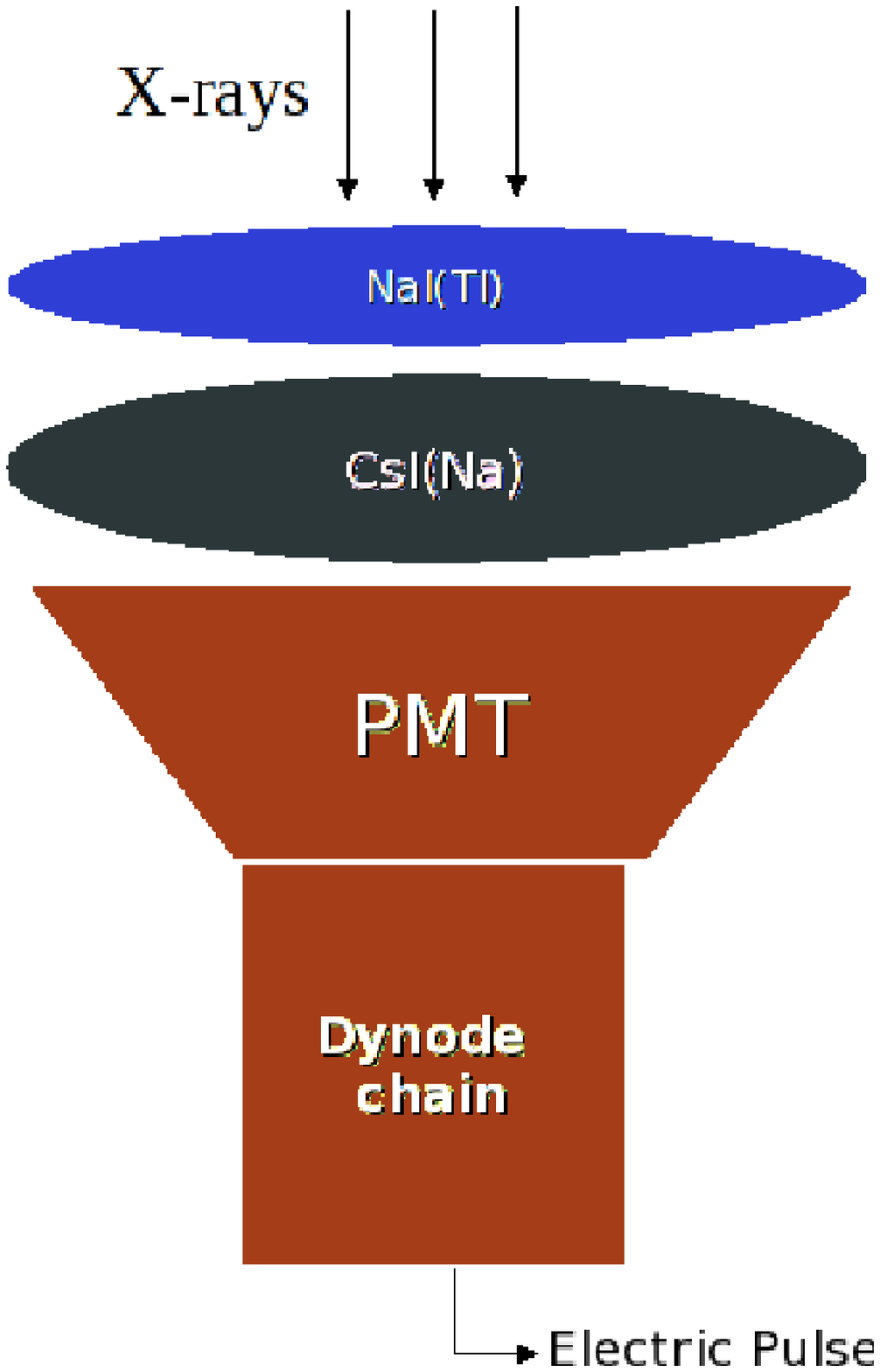}
\hskip 0.5cm
\includegraphics[width=5.5cm]{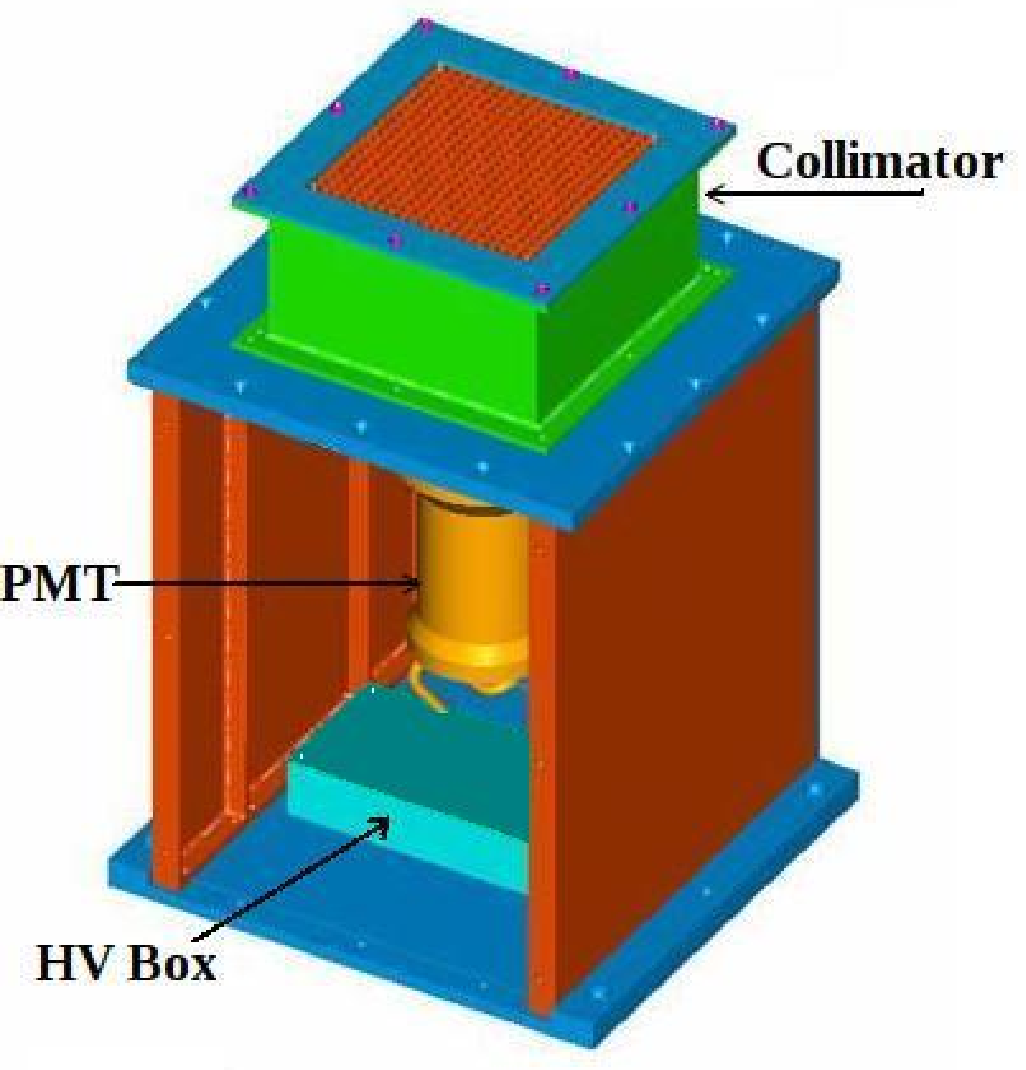}}}
\caption{A Schematic drawing of Phoswich assembly (NaI and CsI crystals) along with the PMT is
shown in the left and a schematic drawing of one of the payloads (RT-2/S) is shown in the 
right.}
\label{kn : fig1}
\end {figure}

The interaction of X-ray photons of energy up to 100 keV with NaI and CsI crystals is fully 
dominated by the photoelectric process and thus absorbed radiation (secondary electron-hole 
pair absorbed by the impurities) converts into light photons (due to the decay of the excited 
impurities). These light photons eventually strike the photo-cathode of the PMT (Knoll 1999) 
(gain $\sim$ 10$^6$) and converted into narrow electrical pulse whose magnitude (pulse height) is 
proportional to the energy of the incident radiation. 

The energy resolution (FWHM) of the scintillator Phoswich is expected to be $18\%$ at $60$~keV 
and the pulse height across the crystal should be less than 3\%. The radioactive 
isotopes $^{241}$Am ($59.5$~keV), $^{57}$Co ($122$~keV), $^{109}$Cd ($22$~keV and $88$~keV) 
were used for laboratory calibration (energies in the parentheses correspond to the major
emission lines of the respective radio-active sources), and results are discussed in 
section \S 4. The entire system will be calibrated in flight using a $100$~nano-Curie $^{57}$Co 
source. 

Electronics of both instruments (schematic drawing of one of the payloads is shown in Figure 1) 
are identical, the only difference is that RT-2/G has one 
thin Al sheet ($\sim$ 2 mm) above its collimator, to strongly decrease the low energy photons in the 
energy range of $13$ to $20$ keV. Indeed, solar flares exhibit a wide 
range in intensity and spectral shape. A bright solar flare with a 
steep spectrum would produce copious amount of low energy (10 -- 20 keV) photons rendering 
spectral measurements above 20 keV difficult as the instrument will be saturated. To deal with such 
contingencies, we covered RT-2/G (identical in all other aspects with RT-2/S) with a filter to 
decrease low energy photons.                                                                                                   

\subsection {Electronics of RT-2/S \& RT-2/G}

Both payloads (RT-2/S \& RT-2/G) are interfaced with the satellite system (SSRNI, a Russian
acronym which means System of Collection and Registration of Scientific Information or SCRSI
in English) through the RT-2 electronic processing device (RT-2/E). RT-2/E receives necessary 
commands from 
the satellite system and passes them to the individual detector systems for proper functioning
of the detector units and it also acquires data from the detector systems and stores in its 
memory for further processing. Detailed description of the functioning of RT-2/E and data 
management scheme are discussed in Sreekumar et al. (2010).

The electronics system in the detector module consists of i) front end electronics for the 
Phoswich scintillator, ii) electronics for pulse shape (PS) and pulse height (PH), iii) 
FPGA (Field Programmable Gate Array) based data packaging system, iv) high voltage (HV) DC-DC 
converters and distribution box and v) low voltage DC-DC converter. The block composition of 
detector electronics is shown in Figure 2.
 
\begin {figure}[h]
\vbox{
\vskip 0.0 cm
\centering{
\includegraphics[width=10.0cm]{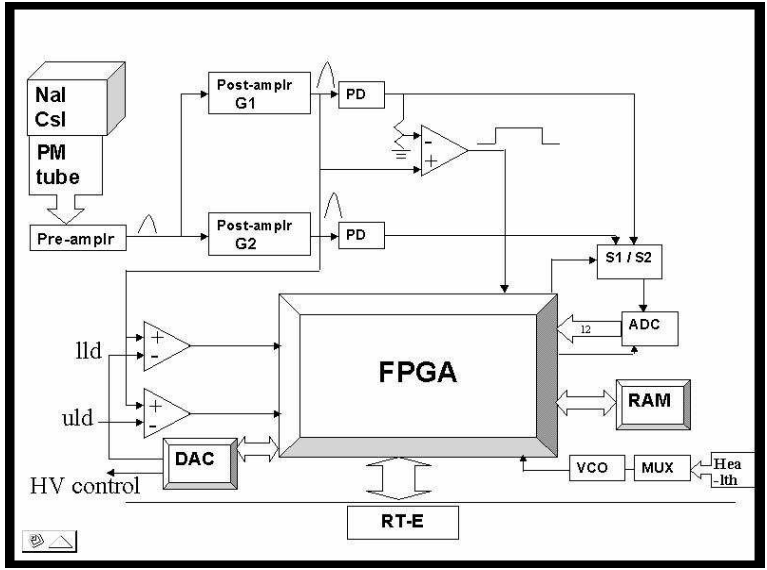}}}
\vskip 0.0cm
\caption{Block diagram of electronic components of RT-2/S and RT-2/G.}
\label{kn : fig2}
\end {figure}

The electronics system receives signals from the NaI(Tl) / CsI(Na) scintillator Phoswich 
detector assembly. These signals are analyzed and the details are discussed in the next section 
(\S 3.3).
The data are encoded in the proper format for dispatching to the RT-2/E device. Since it is very 
difficult to handle a large range of energy by a single amplifier, the entire energy range of the 
Phoswich detector is covered by two different amplifiers, namely G1 and G2. This is done 
electronically by comparing the detector pulse output to two analog voltages LLD (Lower Level
Discriminator) and ULD (Upper Level Discriminator) so 
that pulses between LLD and ULD are deemed to be G1 and above ULD is deemed to be G2. G1 contains 
information pertaining to both NaI(Tl)  and CsI(Na) crystals (which can be segregated using 
the pulse shape) and G2 is presumed to be coming from the CsI(Na) crystal. The pulse shape 
(PS) is measured using a 7 bit counter and the discrimination is done digitally by comparing the 
pulse shape with ps$\_$cut (ps $<$ ps$\_$cut is deemed to be coming from the NaI crystal and 
ps $>$ ps$\_$cut from CsI). 

The pre-amplifier is designed using discrete components. It is used in an inverting operational
amplifier mode which converts the -ve input from PMT into +ve pulse with a gain factor of 1 to 2.
Two post amplifiers (G1 \& G2) with different gains and different filters are used to cover the
different energy ranges of interest. These discrete post amplifiers are of non-inverting nature
with provision of adjustable gain (during ground testing, not by command).
Amplifier G1 will cover energy range from $15 - 210$~keV (the approximate energy ranges are 
15 keV - 100 keV for G1-NaI; 30 keV - 210 keV for G1-CsI) \& G2 from $40$~keV to $1$~MeV. These
ranges can be scaled by changing the HV through command and the LLD (15 keV) can be separately changed.

The NaI and CsI crystals have different decay times. The gain from these two detectors (crystals) 
is a function of the time constant of the pre-amplifiers (which is common for both 
detectors), resulting in a gain difference (of a factor of 1.8) of
the outputs of NaI and CsI for the same energy deposition. These energy ranges were measured 
during the ground calibration. Also G1-NaI, G1-CsI \& G2 spectra and light curves are divided 
into 8 different counters (4 counters for G1-NaI, 2 counters for G1-CsI and 2 counters for G2). 
The channel numbers of these counters and the corresponding energy ranges are shown in Table 1, 
for the default ground settings. These values can be changed by command. The electronics of both 
instruments are identical and have the same working energy ranges. The basic differences of both 
payloads are mentioned in Section 1 and Section 3.1.


\begin{table}[h]
\centering
\centerline {Table 1: RT-2/S \& RT-2/G channel boundary (default) and energy ranges}
\vskip 0.2cm
\begin{tabular}{|l|c|c|c|c|}
\hline
Amplifiers & Counters & Channel ranges &\multicolumn{2}{|c|}{Energy Ranges (keV)*}\\
\cline{4-5}
 & &(default)&~~~~~RT-2/S~~~~~&RT-2/G\\
\hline
        & C1 & 0 - 140   & $<$ 13  & $<$ 13 \\
        & C2 & 141 - 280 & 13 - 27 & 13 - 27\\
G1-NaI  & C3 & 281 - 560 & 27 - 55 & 27 - 56\\
        & C4 & 561 - 1023& 55 - 102& 56 - 104\\
\hline
        & C5 & 0 - 256   & $<$ 33   & $<$ 34\\
G1-CsI  & C6 & 257 - 1023& 33 - 209 & 34 - 209\\
\hline
        & C7 & 0 - 64    & $<$ 48& $<$ 246\\
G2      & C8 & 65 - 255  & 48 - 570& 246 - 1000\\
\hline
\end{tabular}

*Differences in energy boundaries of all 8 counters (C1,...,C8) of both
payloads (RT-2/S and RT-2/G) are due to different technical setting of operational
conditions (eg. HV, LLD, ULD, PS cut etc.).
\end{table}

The operational aspects of both payloads are identical. They operate with 27$^{+7}_{-3}$ 
Volt (Power bus of 27$^{+7}_{-3}$ V, from the satellite system, is routed through RT-2/E). The 
total power consumption is limited to 4.5$\pm$0.5 Watt for each payload. 
A low voltage MDI DC-DC converter is used to convert the input voltage (27 Volt) to $\pm$ 15 Volt 
and +5 Volt to drive the different components of the individual payloads.
The +15 Volt is also converted separately through a voltage regulator circuit to 
high voltage ($\sim$ 700 Volt) needed for biasing the PMT that is operable in a
400-900 V range. The HV can be commanded to any value in this range with an 
accuracy of $\sim$ 4.5 Volt. The Pulse Shape Discriminator (PSD) of the two crystals (NaI \& CsI) 
and Lower Level Discriminator (LLD) for the two amplifiers (G1 \& G2) can be also commanded. 
Voltage Control Oscillator (VCO) is used to monitor the instrument health parameters, such as 
+5 Volt supply, temperature (Thermistor), High Voltage (HV) \& LLD.

\subsection {Signal processing method in RT-2/S \& RT-2/G}

When an incident radiation (X-ray event) interacts with the NaI(TI) or CsI(Na) crystal, it 
gives out light photons which are converted into electrical pulses by a PMT. Light outputs 
from NaI and CsI crystals have decay time scales of 250 ns and 650 ns respectively. Pulses 
from the detector (crystal) are amplified in a pre-amplifier and two post amplifiers (G1 and G2). 
The output from the selected post amplifier (G1 or G2) is digitized using an analog to digital 
converter (ADC). The ADC is 12 bits wide and has a typical conversion time of 10 $\mu$s. The block
diagram of time scale for processing of an event is shown in Figure 3 and the
events have a fixed dead time of $\sim$20 $\mu$s.

\begin {figure}[h]
\vbox{
\vskip 0.0 cm
\centering{
\includegraphics[width=10.0cm]{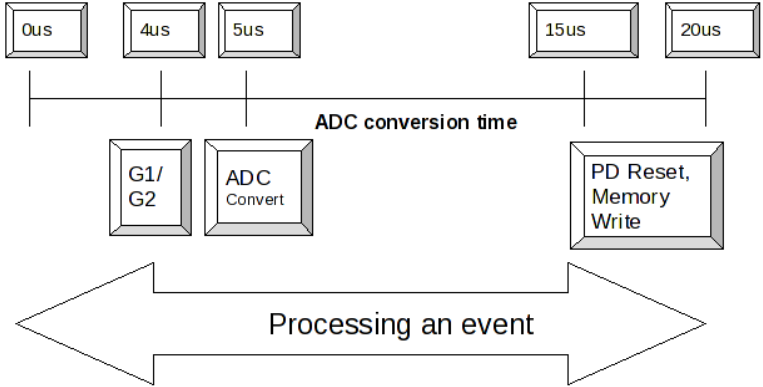}}}
\vskip 0.0cm
\caption{Block diagram of the time scale for processing of an event.}
\label{kn : fig3}
\end {figure}

The peak of the pulse output from the detector is proportional to the charge deposition in the 
crystal and hence the pulse height (PH) denotes the radiation energy. PH is converted into a 12 bit 
digital data and sent to FPGA. PS denotes the pulse decay time,
which gives a hint of whether the event is occurring in NaI or CsI. PSD 
data is directly transmitted to the FPGA in a 1 bit format. The value of this bit (i.e, based on 
the PS information) is determined by the FPGA. This data is stored, by the FPGA, into a memory, 
for a time determined by RT-2/E. RT-2/E will read this data every second. In the Normal mode, the 
time required to send the data from RT-2/S (RT-2/G) to RT-2/E is $\sim0.21$ sec. In the Event mode, 
it depends on the number of events occurred and the maximum time required is 1 sec. The maximum 
number of events to be handled in event mode is 7360 events. This restriction is due to the 
onboard memory size. Detail discussion of the two operational modes (Normal and Event mode) are 
given in the next section. 


	
\subsection {Data specifications and operation of RT-2/S \& RT-2/G}

As we already mentioned, there are two different modes in which RT-2/S and RT-2/G can 
be operated - the Event mode and the Normal mode. In the Event mode, every event is 
time-tagged with an time accuracy to 0.3 ms and in the Normal mode all the spectral data are 
accumulated. Here, eight channel count rates are stored every 10 ms and are sent to RT-2/E every 
second. The Normal mode data are processed in RT-2/E with a best possible time 
resolution of 1 s for the full spectrum and 10 ms for the 8 channel count rates.

A `pure' event (registered X-ray event on the detector) consists of a pulse height (PH) of 12 bits, 
a pulse shape (PS) of 7 bit, an amplifier identification (G1 or G2) of 1 bit and the time of 
registration of the events of 12 bits. Therefore, the first 20 bits carry the spectral information 
and the last 12 bits carry the time-tagged information of each event. In the Event mode, time-tagged 
information is associated with the events and it is incremented after every event, whereas in the 
Normal mode time-tagged information is used to count the total no. of events that are registered in 
every 10 ms (calculation is done internally by the FPGA of the detector processing electronics). 
Data specifications of the two different modes are given below.

\subsubsection {Event mode}

In the Event mode, each detector `event' is coded into 32 bits (4 bytes). The maximum number of 
events that can be stored in the memory is 7360 events, which are sent to RT-2/E, whenever `1 sec' 
command is received. The event data structure is given in the following format:

\begin{itemize}

\item[$\bullet$]{\bf Event block:} 32 bits $\times$ number of events (the number of maximum events 
that can be stored in the memory is 7360). Each event consists of 32 bits {\it (D0-D11: 12 bit ADC, 
D12: 1 bit G1/G2 Selection, D13-D19: 7 bit PSD and D20-D31: 12 bit Timing)} data. 

\end{itemize}
 
The Event mode is basically used to verify the detector functionality as in this mode, the data 
directly come to the processing electronic device (RT-2/E) from the detectors. For scientific 
purposes, this mode could be operated to make high time resolution (0.3 ms) observations such as 
the Crab Nebula. Due to the onboard memory restriction, `memory full signal' will be noticed for 
an observation longer than $\sim 200$s and the system will go to the `BAD' mode 
(i.e., non accumulation of scientific data). Restriction of 
the operation in this mode could be done by time-tagged commands to avoid memory overflow.

\subsubsection {Normal mode}

In the Normal mode, the event data (20 bits of scientific information and 12 bits of timing data), 
along with the health parameters encoded in VCO, are packaged by the FPGA and kept in a memory 
page. They contain a header block including the health parameters and the counts, spectrum block, timing 
block with high time resolution of the counter. The normal mode data has the following format:

\begin{itemize}

\item[{\large $\bullet$}] {\bf Header block (18 bytes):} VCO (2 bytes) and 
8 counters (16 bits each) based on the PH values. The counter values are total 
counts over 4 channel intervals (PH between 0 to ch1, ch1 to ch2, ch2 to ch3 
and $>$ ch3) in G1-NaI; over 2 channel intervals (0 to ch5 and $>$ ch5) in G1-CsI 
and over 2 channel intervals (0 to ch7 and $>$ ch7) in G2. 
Header contains mode ID - 0: Normal mode and 1: Event mode.

\item[{\large $\bullet$}] {\bf Spectrum block (4864 bytes):} G1-NaI (PH) : 
$1024 \times 2$ bytes + G1-CsI (PH) : $1024 \times 2$ bytes + G2 (PH) : $256 \times 2$ 
bytes + PS : $128 \times 2$ bytes.

\item[{$\bullet$}] {\bf Timing block (1600 bytes):} 100 Timing blocks 
$\times 8$ counters $\times 2$ bytes (counters in each bin will count for 
10 ms).

\end{itemize}

\begin{table}[h]
\centering
\centerline {Table 2: RT-2/S and RT-2/G data specifications}
\vskip 0.2cm
\begin{tabular}{|l|c|c|c|}
\hline
\multicolumn{4}{|l|}{\bf VCO data: (2 bytes):}\\
\hline
D15 &\multicolumn{2}{|c|}{D14-D12} & D11-D0 \\
\hline
Mode Id (0/1) &\multicolumn{2}{|c|}{VCO Channel Numbers (0-7)} & VCO Counts (values) \\
\hline
\hline
\multicolumn{4}{|l|}{\bf Scientific Data in the NORMAL Mode (Id: `0') (Total 6480 bytes):}\\
\hline
         & G1-NaI & 2048 bytes & 1024 spectral ch. $\times$ 1 word*\\
\cline{2-4}
Spectrum & G1-CsI & 2048 bytes & 1024 spectral ch. $\times$ 1 word\\
\cline{2-4}
         & G2     & 512 bytes & 256 spectral ch. $\times$ 1 word\\
\cline{2-4}
         & PSD    & 256 bytes & 128 spectral ch. $\times$ 1 word\\
\hline
Timing & 800 words & 1600 bytes & 8 ch. $\times$ 100 blocks $\times$ 1 word\\
\hline
Counters & 8 words & 16 bytes & 8 counters $\times$ 1 word\\
\hline
\hline
\multicolumn{4}{|l|}{\bf Scientific Data in the EVENT Mode (Id: `1') (4 bytes/event):}\\
\hline
Time & PSD & G1/G2 & ADC \\
\hline
D31-D20 & D19-D13 & D12 & D11-D0 \\
\hline
\end{tabular}

*1 word = 2 bytes = 16 bits.
\end{table}

The data structure of stored (scientific) data of RT-2/S and RT-2/G are summarized in Table 2.
On the receipt of a `1 sec' command from RT-2/E (issued every second), the stored data are sent 
to RT-2/E for further processing in specific formats (Sreekumar et al. 2010).
The header, the spectral and the timing blocks are sent in the Normal mode and 
the header and the event block are sent in the Event mode. These modes of operation could be 
decided by ground commands (Sreekumar et al. 2010).




\section{Tests and Evaluation of the RT-2/S \& RT-2/G payloads} 

It is essential to carry out extensive space qualification tests (on ground)
of the scientific payloads in order to ensure that they can sustain and perform
according to expectations during their in-flight lifetime.
An instrument's proper functionality (scientific performance) is based on 
High Voltage (HV) and LLD/ULD operation, stability of Pulse shape (PS) and Pulse height (PH),
health parameters, quality of spectral (scientific) data etc. Both payloads (RT-2/S and
RT-2/G) were space qualified independently. 
The space qualification tests we performed include:

\begin{itemize}

\item[{$\bullet$}] {\bf Vacuum Test:} Payloads were kept in Vacuum chamber of pressure 
$\sim$10$^{-6}$ mm of Hg at 23$^\circ$C. During this test, the instrument's functionality was 
verified. Major test result of HV variation is summarized in Section 4.3.

\item[{$\bullet$}] {\bf Thermal Test:} Both payloads were tested at various temperature 
conditions ranging from -10$^\circ$C to 40$^\circ$C. In this condition, the functional tests 
results are summarized in Section 4.4 and 4.5.

\item[{$\bullet$}] {\bf Vibration Test:} Vibration test is essential to check the resonance effect
on the detector structure as well as electrical card systems of the payloads. Before and after
this test, instruments functionality were verified and we did not find any malfunction in both
instruments.

\item[{$\bullet$}] {\bf EMI Test:} The electro-magnetic interference (EMI) test is important for the 
RT-2 instruments. Radiation from the other devices (such as radio transmitting device on the 
spacecraft) may affect the normal behavior of the RT-2 payloads. While doing this test, 
functionality of both detectors were verified.

\item [{$\bullet$}] {\bf SRC Test:} The standard room condition (SRC) test was carried to check 
the normal functionality of both detectors. This test was carried out after each space 
environment test. This test includes the verification of total power consumption, health 
parameter, HV and LLD/ULD operation, scientific data etc.

\end{itemize}

During these laboratory tests, the complete functioning of both RT-2/S and
RT-2/G payloads were investigated and we calibrated HV, LLD, PS and PH
variation with temperature, energy spectra etc. Testing
of both payloads were carried out for a period of three months.
The results are discussed in the following sections. In the next section, the lab testing set-up 
is described.

\subsection{Testing set-up of RT-2/S and RT-2/G}

The overall testing of RT-2/S, RT-2/G, RT-2/CZT \& RT-2/E were carried out with Ground Check-Out 
system (KIA, a Russian acronym). In reality, KIA mimics the total functionality of the Satellite 
system (SSRNI) for the RT-2 system. We used the check-out software written in LabVIEW platform 
to test the individual and complete packages (payloads).
Details of the overall set-up schemes are discussed in Sreekumar et al. (2010).
 

We tested the instruments by directly interfacing it with a PC via the OPTO device, the SCB-68 
connector box and the NI data acquisition card (PCI 6534). The OPTO device was used to 
isolate the payloads electrically from the PC system using a buffer and an opto-isolator. The 
LabVIEW based software along with the OPTO device mimics the functioning of the processing 
electronic device (RT-2/E). Figure 4 shows the schematic block diagram of the testing set-up used.

\begin {figure}[h]
\vbox{
\vskip 0.0 cm
\centering{
\includegraphics[width=7.0cm,angle=270]{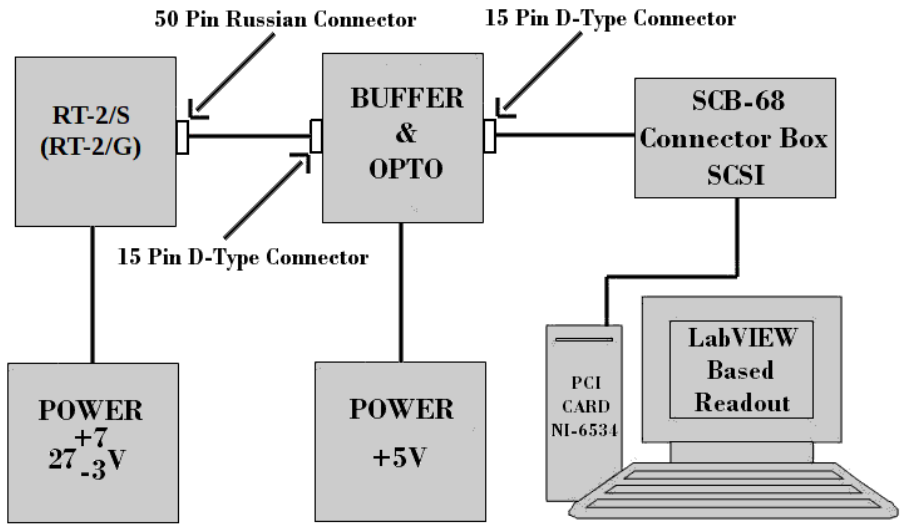}}}
\vskip 0.0cm
\caption{Schematic block diagram of laboratory set-up for testing of payloads.}
\label{kn : fig4}
\end {figure}

The complete control of the instrument operation was done by executing series of commands 
setting the channel boundary, the HV, the PS cut and the LLD value of each payload. The 
command structure used to set the different instrument parameters is given in Table 3.

RT-2/S \& RT-2/G are normally operated with a high voltage (HV) of $\sim$ 710 Volt (command: 
hx07A6 for S, hx27A6 for G), a LLD threshold of 0.86 V (command: hx0628 for S, hx2628 for G) 
and a PS cut value equal to channel 26 (command: hx051A for S) and channel 28 (command: hx251C 
for G). 

\begin{table}[h]
\centering
\centerline {Table 3: RT-2/S \& RT-2/G detector commands}
\vskip 0.2cm
\begin{tabular}{|c|c|c|l|}
\hline
{\bf Command Type}&\multicolumn{2}{|c|}{\bf HEX Command of}&{\bf Description}\\
\cline{2-3}
 & RT-2/S & RT-2/G & \\
\hline
`1 sec' & 0800 & 2800 & Set detector in normal mode\\
\cline{2-4}
command & 0801 & 2801 & Set detector in event mode\\
\cline{2-4}
 & 0802 & 2802 & Corona override*\\
\hline
Range setting & 00xx & 20xx & Set ch. boundary for G1-NaI (ch1)\\
\cline{2-4}
command**  & 01xx & 21xx & Set ch. boundary for G1-NaI (ch2)\\
\cline{2-4}
(value = `xx' times 4, & 02xx & 22xx & Set ch. boundary for G1-NaI (ch3)\\
\cline{2-4}
except ch7 boundary) & 03xx & 23xx & Set ch. boundary for G1-CsI (ch5)\\
\cline{2-4}
	 & 04xx & 24xx & Set ch. boundary for G2 (ch7)\\
\hline
PS cut value &05xx & 25xx& Set PS cut value (ps$\_$cut) to separate\\
(7bit value) &     &     & G1-NaI \& G1-CsI spectra\\
\hline
DAC setting & 06xx & 26xx & Set LLD value from DAC\\
\cline{2-4}
(8bit value)& 07xx & 27xx & Set HV value from DAC\\
\hline
RAM select & 0c00 & 2c00 & Select section 1 (256 Kbytes) of RAM\\
\cline{2-4}
command & 0e00 & 2e00 & Select section 2 (256 Kbytes) of RAM\\
\hline
\end{tabular}

*A facility is also kept to suppress the automatic corona sensing and this can be activated by 
the `corona-override' command.

**Range setting command: `xx' refers to HEX value. The actual channel boundary is given by four
times of the `xx' value for ch1, ch2, ch3 and ch5, whereas for ch7 the value is just `xx'. For 
example, if channel boundary of `ch1' is to be set at 140 channel, then `xx' value should be hx23 
(as hx23 = dx35 and original channel value is 35 x 4 = 140 channel).

\end{table}

\subsection{Health parameters of the RT-2/S \& RT-2/G instruments}

There are eight VCO channels giving health information on each instrument. 
Six over the eight channels were configured to certain voltage levels, which are given in 
Table 4. Each VCO channel shows some count values that can be related to a physical or 
instrumental parameter thanks to a dedicated calibration. As an example, the variation of VCO 
counts for HV variation is shown in Figure 5. 
The different health parameters to be monitored during various space environmental tests are 
given in Table 4. 
The channels 3 \& 7 were kept unused.

\begin{table}[h]
\centering
\centerline {Table 4: RT-2/S \& RT-2/G health parameters (VCO channels)}
\vskip 0.2cm
\begin{tabular}{|c|c|c|}
\hline
{\bf Channel No.}&{\bf Description}&{\bf Operating Voltage}\\
 & & {\bf Level}\\
\hline
0 & Supply Voltage & 5.0 $\pm$ 0.5 V\\
\hline
1 & Thermistor & 1.5 - 5.5 V\\
\hline
2 & Supply Analog & 5.0 $\pm$ 0.5 V\\
\hline
3 & NC* &  0 V\\
\hline
4 & HV Feedback & 1.5 - 5.0 V\\
\hline
5 & HV Reference & 1.5 - 5.0 V\\
\hline
6 & LLD & 0.5 -1.6 V\\
\hline
7 & NC* &  0 V\\
\hline
\end{tabular}

*NC: not configured
\end{table}

\subsection{High Voltage (HV) calibration of RT-2/S}

The normal operation of HV in space environment is an important issue, as the present detectors
functionality depends on HV operation. So, we calibrated the HV of both instruments during the
Vacuum test, which mimics the space environment conditions.   
The high voltage (HV) calibration of the RT-2/S payload was done by changing input digital to analog
converter (DAC)
values by steps through commands. Each DAC value corresponds to different HV and so different 
VCO counts. DAC values were varied from 134 to 208 and corresponding VCO counts (channel 5) were 
found to be varying from 4976 to 11280. The HV calibration result is shown in Figure 5. It is 
found that the instrument is operable in the linear regime of HV variation from 565 to 900 Volt
(within the specified operating range of PMT).
During each step of HV variation, we checked the Pulse height (PH) variation, as it is important
for calibration of the energy spectrum. 
We also found similar behavior of HV variation of the RT-2/G payload. 

\begin {figure}[h]
\vbox{
\vskip 0.0 cm
\centering{
\includegraphics[width=5.0cm,angle=270]{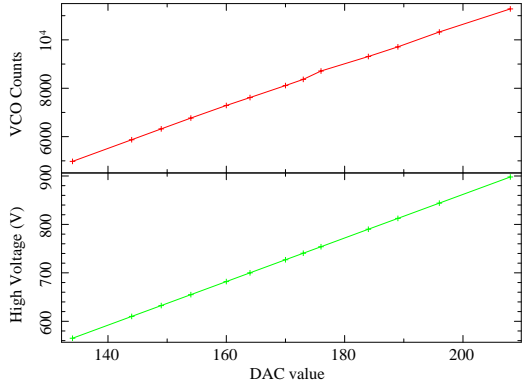}}}
\vskip 0.0cm
\caption{High Voltage (HV) calibration with DAC value (commands).}
\label{kn : fig5}
\end {figure}

\subsection{Thermistor calibration of RT-2/S}

The thermistor is the temperature sensor placed inside each instrument to monitor its 
temperature variation. The sensor (resistor-type) is the same on both instruments. The thermistor 
resistance, which changes with temperature, is measured as VCO counts. Through the 
monitoring of the VCO counts (channel 1), we can calibrate the temperature of the payload. 
The instrument is operable in the temperature range from -10$^\circ$C to +40$^\circ$C. 
Therefore, we measured the temperature variation of the thermistor in the same temperature 
range. The VCO count variation with temperature is shown in Figure 6. The plot is not a
linear function, because the in-built resistance of the thermistor varies non-linearly
with temperature. The thermistor resistance is coupled to the VCO circuit to measure the
voltage change (see Table 4) and hence the VCO counts. Looking at the VCO counts 
(channel 1) from the scientific data, one can estimate the onboard payload temperature. 

We did the temperature related test of both instruments for long duration. We kept the 
instruments inside a thermal chamber for at least 3-4 hrs to attain a particular temperature, 
such that each part of the instrument get heated uniformly. Once the thermal equilibrium is 
reached, we verified the normal functionality (health parameter, temperature, spectral 
information etc.) of each instrument.
We also found similar behavior in temperature variation for the RT-2/G payload. 

\begin {figure}[h]
\vbox{
\vskip 0.0 cm
\centering{
\includegraphics[width=5.0cm,angle=270]{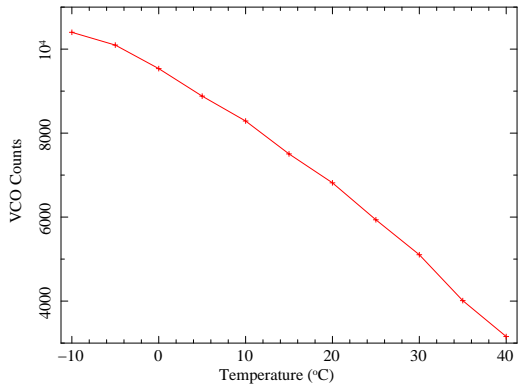}}}
\vskip 0.0cm
\caption{Thermistor calibration: Variation of VCO values with temperature for RT-2/S payload.}
\label{kn : fig6}
\end {figure}

The nominal operating temperature of both instruments is $\sim$23$^\circ$C. 
Initial results from the onboard data showed that the temperature of both payloads are
maintained at comfortable operating range of +18$^\circ$C to +25$^\circ$C (Nandi et al. 2009).

\subsection{Pulse shape (PS) \& Pulse height (PH) variation with temperature of RT-2/S}

The study of PS and PH variation is important as both change with temperature
and therefore this may have some consequences on the operation of the
instruments. Indeed, PS denotes whether an event is occurring in the NaI or
CsI crystal in the PSD spectrum. Therefore, the stability of PS is crucial to
ensure the reability of the spectral information coming from both crystals.
So, if there is any change in the PS cut value due to a temperature change, we
can adjust the change in PS cut through commands. PH gives some information
on the energy of an event. As an example, PH has a value of around channel 600
for $^{241}$Am 59.5 keV photons in the NaI crystal. From channel-energy
calibration (discussed in the next section), we can derive the energy
resolution as well as the gain (keV/Chan) of both instruments.  If
there is any change in gain, we can adjust the gain through a HV command.

We made a systematic study of PS and PH variation with temperature over the
in-flight operating temperature range (from -10$^\circ$C to +40$^\circ$C). In Figure 7(a), the
PS variation with temperature is shown. PS appears to be almost stable over
the operating temperature range with a maximum variation of 7\% with respect to
the average PS value.

\begin {figure}[h]
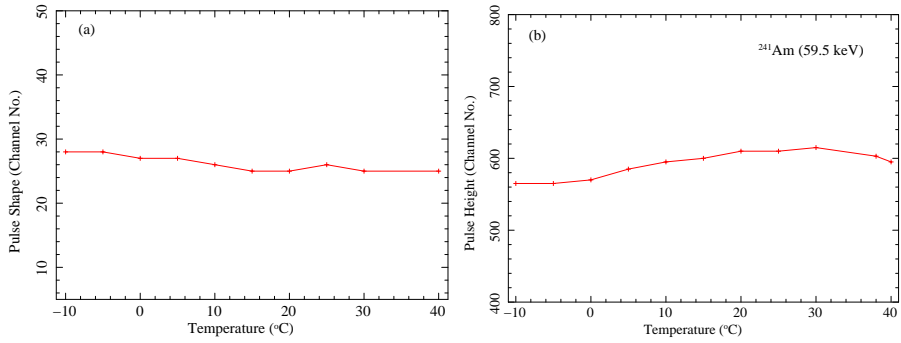

\vbox{
\vskip 0.0 cm
\centering{
\includegraphics[width=4.5cm,angle=270]{fig7a.ps}
\hskip 0.02 cm
\includegraphics[width=4.5cm,angle=270]{fig7b.ps}}}
\vskip 0.0cm
\caption{(a) Pulse shape (PS) variation as a function of temperature and (b) Pulse Height (PH) 
variation ($^{241}$Am source peak @ 59.5 keV) of G1-NaI spectrum with temperature.}
\label{kn : fig7(a-b)}
\end {figure}

In Figure 7(b), we plotted the pulse height (PH) variation of G1-NaI spectra with temperature 
varying from -10$^\circ$C to 40$^\circ$C. To study the temperature effect on PH, we used the 
$^{241}$Am 59.5 keV line as a reference. At the room temperature (+$25^\circ$C), PH is stable at 
around channel 600. The PH variation is observed to be less than 6\% over the entire temperature 
range.

We also performed the same PH measurement for G1-CsI spectrum and found similar PH variation
using a $^{57}$Co 122 keV line in the G1-CsI spectrum. But, we did not measure the PH
variation of G2 spectra. PS measurement is not applicable to G2 spectrum, as G2 spectral data
directly come from the CsI crystal.

We also thoroughly studied the PS and PH variation of RT-2/G detector with temperature and we 
found similar results for PS and PH variation of RT-2/G.

\subsection{Calibration of the G1-NaI spectral response of both RT-2/S and RT-2/G}

To calibrate the G1-NaI spectrum (of both instruments), which covers 1024 channels, we used 
two strong radio-active sources ($^{241}$Am \& $^{109}$Cd). The main emission line energies of 
$^{241}$Am are at 13.95 (13.3\%), 17.74 (19.4\%), 20.8 (4.9\%), 26.35 (2.4\%) \& 59.54 (35.8\%) keV 
and for $^{109}$Cd are at 22.16 (84\%) \& 88.04 (4\%) keV (in the parenthesis, the source strength of
respective emission lines is mentioned. See for reference, http://ie.lbl.gov/education/isotopes.htm). 
For the calibration purpose, we used only three emission peaks 
(59.54 keV of $^{241}$Am and 22.16, 88.04 keV of $^{109}$Cd) of these two sources, as all of
the three lines are well separated in energies (no chance of blending with one another). The low
energy emission lines (13.95 and 17.74 keV) are not used for calibration, as both lines are 
not resolved completely in RT-2/S and suppressed in RT-2/G due to the 
Al filter and too weak when compared to the 22 keV emission line of $^{109}$Cd emission. 
The energy spectra of G1-NaI of both instruments are shown in Figures 8(a,b) and 9(a,b). Both
spectra are taken in identical operational condition with the same exposure time of 200 sec. The
standard emission lines of $^{241}$Am and $^{109}$Cd are detected in the NaI spectra along with
a weak background continuum. The difference in peak amplitude of $^{241}$Am (or $^{109}$Cd) 
emission feature in RT-2/S and RT-2/G is mostly due to the difference in operational 
setting conditions, such as HV and LLD setting, PS cut value etc. of both instruments. The 
significant amplitude difference could not be due to Quantum Efficiency (QE) difference (although 
we did not perform any QE test for both instruments) as both the crystals (detectors) are 
completely identical in nature.

\begin {figure}[h]
\vbox{
\vskip 0.0 cm
\centering{
\includegraphics[width=4.5cm,height=5.0cm,angle=270]{fig8a.ps}
\hskip 0.01cm
\includegraphics[width=4.5cm,height=5.0cm,angle=270]{fig8b.ps}
\vskip 0.1cm
\includegraphics[width=5.5cm,height=5.0cm,angle=0]{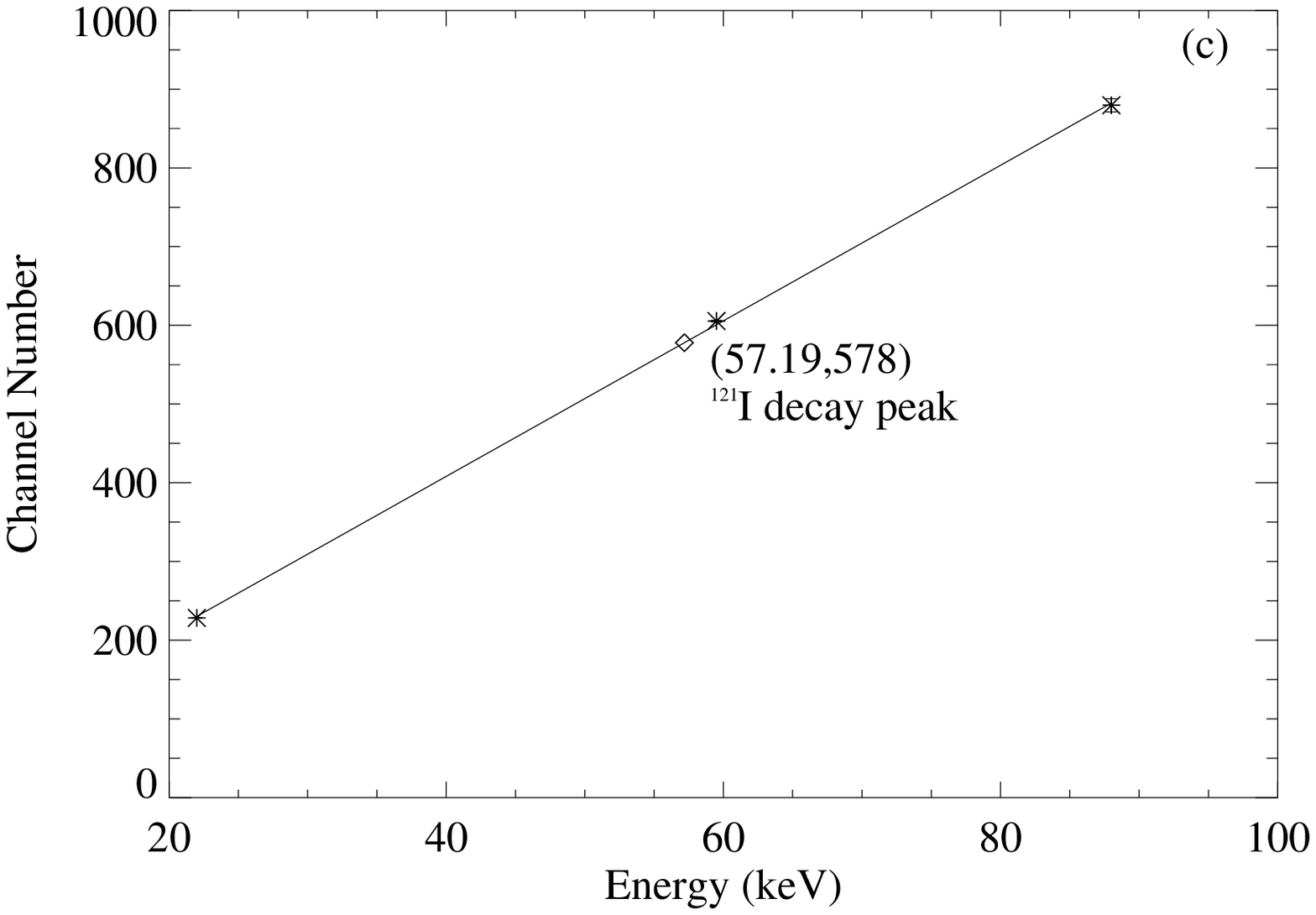}}}
\hskip 0.01cm
\caption{Energy calibration of RT-2/S. (a) The top left panel shows the G1-NaI 
spectrum obtained using a $^{241}$Am source. Note that the 13.94 \& 17.74 keV lines and 20.8 \& 
26.35 keV lines are blended; (b) the right panel shows the G1-NaI spectrum obtained using 
$^{109}$Cd (emission peaks: 22 keV, \& 88 keV); (c) bottom panel: the channel-energy 
calibration curve is plotted with the mark of identification (diamond point) of the 
$^{121}$I decay peak at 57.19 keV (see text for details).}
\label{kn : fig8(a-c)}
\end {figure}

We fitted each peak profile (marked with solid black line) using a Gaussian in order to find their 
centroid values (channel
numbers). Importantly, it is noted that the $^{241}$Am (59.5 keV) line is asymmetrical in nature
towards the low energy (Figures 8(a) and 9(a)), while other emission lines are almost symmetrical 
in nature. The asymmetry is due to the superposition of the background line ($^{121}$I decay peak, 
channel values are close to the $^{241}$Am @ 59.5 keV line, see Table 5). This line is clearly 
seen in Figures 8(b) and 9(b). 
The centroid values of all the emission lines are measured in channel number
using the background subtracted data.
The characteristics of the background spectrum are discussed in the next section.

\begin {figure}[h]
\vbox{
\vskip 0.0 cm
\centering{
\includegraphics[width=4.5cm,height=5.0cm,angle=270]{fig9a.ps}
\hskip 0.01cm
\includegraphics[width=4.5cm,height=5.0cm,angle=270]{fig9b.ps}
\vskip 0.1cm
\includegraphics[width=5.5cm,height=5.0cm,angle=0]{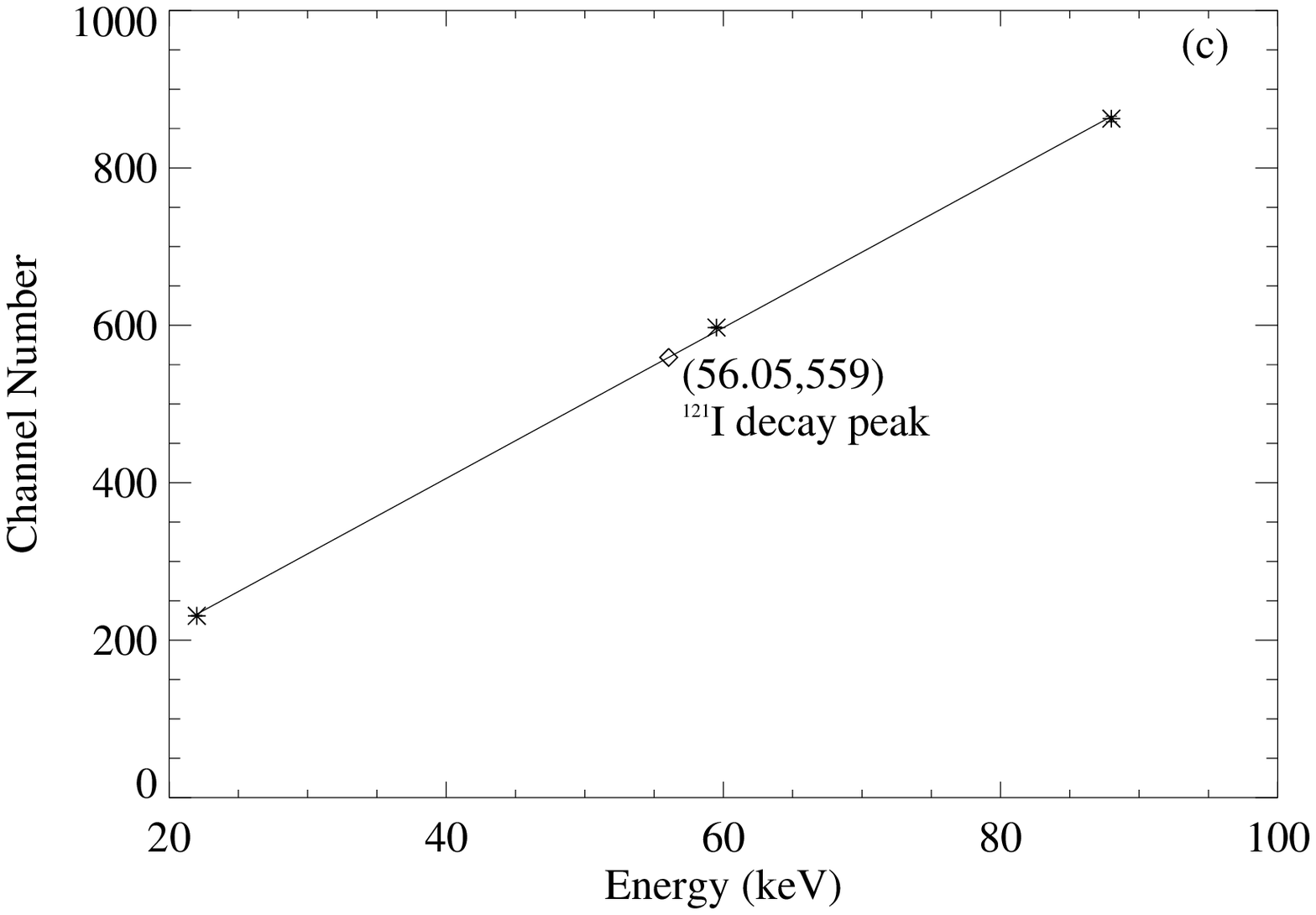}}}
\hskip 0.01cm
\caption{Energy calibration of RT-2/G. (a) The top left panel shows the G1-NaI
spectrum obtained using a $^{241}$Am source. Note that the 20.80 keV and 26.35 keV
lines are blended; (b) the right panel shows the G1-NaI spectrum obtained using
$^{109}$Cd (emission peaks: 22 keV, \& 88 keV); (c) bottom panel: the channel-energy
calibration curve is plotted with the mark of identification (diamond point) of the 
$^{121}$I decay peak at 56.05 keV (see text for details).}
\label{kn : fig9(a-c)}
\end {figure}

Knowing the channel values of the line centroids, we are able to compute the energy scale. 
Since the detector spectral response 
should be linear, we fitted a linear function to our measurements using the least square 
fit technique. The G1-NaI channel-energy calibration plots are shown in Figure 8(c) and in 
Figure 9(c) for both instruments. The derived relation is useful to calibrate the first 
four channel boundaries (see Table 1) of both instruments. The lower and upper energy 
thresholds of the G1-NaI energy scale are around 13 keV and 100 keV with a gain factor 
of 0.101 keV/Chan for RT-2/S and of 0.104 keV/Chan for RT-2/G, respectively. The differences 
in energy boundaries of all four channels between both instruments (see Table 1) are due to 
the different technical settings of the operational conditions (e.g. the HV setting, LLD/ULD 
setting and PS cut values etc.).


Figure 8(a) shows two emission lines at low energy. The two features
correspond to two blended lines: i) 13.95 keV and 17.74 keV; ii) 20.8 keV and
26.35 keV. Because it is impossible to resolve these lines, we used a single
Gaussian line profile to fit these blended lines and to derive their mean line
centroid. Note that for RT-2/G only the second blended line (20.8 keV and
26.35 keV) is seen due to the Al filter. The fitted channel numbers and
calibrated energy values of all the different emission lines are shown in
Table 5 along with their corresponding 1 sigma errors.
From Figures 8(a) and 9(a), it is evident that though the lower energy threshold of both instruments 
are at $\sim$13 keV (see Table 1), there is a significant difference in amplitude at lower 
energies due to the presence of the Al filter.

\begin{table}[h]
\centering
\centerline {Table 5: Identification of emission lines and their channel-energy calibration}
\vskip 0.2cm
\begin{tabular}{|l|c|c|c|c|}
\hline
{\bf Energy Peaks}&\multicolumn{2}{|c|}{\bf RT-2/S}&\multicolumn{2}{|c|}{\bf RT-2/G}\\
 \cline{2-5}
(keV) &{\bf Channel No.}&{\bf Energy (keV)}&{\bf Channel No.}&{\bf Energy (keV)}\\
\hline
\multicolumn{5}{|l|}{\it Emission lines used for calibration:}\\
\hline
& & & & \\
{\bf 22.00 }& $228.22_{-0.32}^{+0.28}$ & $21.796_{-1.033}^{+1.058}$ & $230.95_{-0.35}^{+0.35}$ & $21.779_{-1.118}^{+1.148}$ \\
& & & & \\
\hline
& & & & \\
{\bf 59.50 }& $605.41_{-0.91}^{+0.89}$ & $59.971_{-1.499}^{+1.537}$ & $597.22_{-0.92}^{+0.88}$ & $60.011_{-1.624}^{+1.668}$ \\
& & & & \\
\hline
& & & & \\
{\bf 88.00 }& $879.69_{-8.69}^{+8.21}$ & $87.732_{-1.839}^{+1.885}$ & $862.57_{-3.97}^{+3.13}$ & $87.709_{-1.991}^{+2.045}$ \\
& & & & \\
\hline
\multicolumn{5}{|l|}{\it Calibrated emission lines:}\\
\hline
& & & & \\
{\bf 13.94 \& 17.74}& $183.63_{-0.93}^{+0.87}$ & $17.283_{-0.977}^{+1.002}$ & - - - & - - - \\
{\bf (blended peak)}& & & & \\
\hline
& & & & \\
{\bf 20.8 \& 26.35 }& $274.06_{-3.46}^{+2.94}$ & $26.435_{-1.089}^{+1.116}$ & $282.79_{-1.49}^{+1.81}$ & $27.190_{-1.189}^{+1.221}$ \\
{\bf (blended peak)}& & & & \\
\hline
& & & & \\
{\bf $^{121}$I decay peak }& $577.92_{-4.02}^{+4.08}$ & $57.189_{-1.465}^{+1.502}$ & $559.31_{-2.01}^{+1.89}$ & $56.054_{-1.572}^{+1.614}$ \\
{\bf ($\sim$58.00 keV)}& & & & \\
\hline
\end{tabular}
\end{table}


As mentioned earlier, there is a prominent background line observed at around
58 keV (see Figs. 8(b) \& 9(b)).
This line is due to activation of the Iodine in the NaI/CsI crystal via
inelastic scattering of $^{57}$Co gamma-rays producing a $^{121}$I decay peak (see 
Gruber et al. 1996, Rothshild et al. 1998, Nandi et al. 2009). 
As the exact energy of this emission line is `unknown', we could not use it in the calibration 
of the energy scale even if its intensity is larger than the 88 keV $^{109}$Cd line. Using the 
derived channel-energy relation, the $^{121}$I decay peak line is found to be 
at 57.19 keV and 56.05 keV for RT-2/S and RT-2/G respectively (see Table 5).
However, once the energy of this particular line is known, it becomes an useful tool for
onboard calibration of G1-NaI spectra for both instruments. 

The scintillator crystals are low to medium energy resolution ($\Delta E/E$, where $\Delta E$ is
the FWHM of the Gaussian at the particular energy E) detectors with 
energy resolution of about 18\% at 60 keV. The energy calibration need to be a factor of 10 
better than the basic energy resolution, hence we have targeted an energy calibration accuracy 
of about 1 - 2 keV. The measured G1-NaI energy resolution is around 17.54\%$\pm 0.55$
and 17.68\%$\pm 0.27$ at 59.5 keV for the RT-2/S and RT-2/G instruments, respectively. 
We performed similar energy resolution measurements with other sources to study the behavior 
of energy resolution {\it vs.} energy (Figures 
10(a) and 10(b)). The results are quite satisfactory and both the plots (i.e.,
energy resolution vs energy) are fitted with a simple power-law of 
index -0.123 and -0.127 for NaI spectra of S and G respectively.



\begin {figure}[h]
\vbox{
\vskip 0.0 cm
\centering{
\includegraphics[width=5.2cm,height=4.8cm,angle=0]{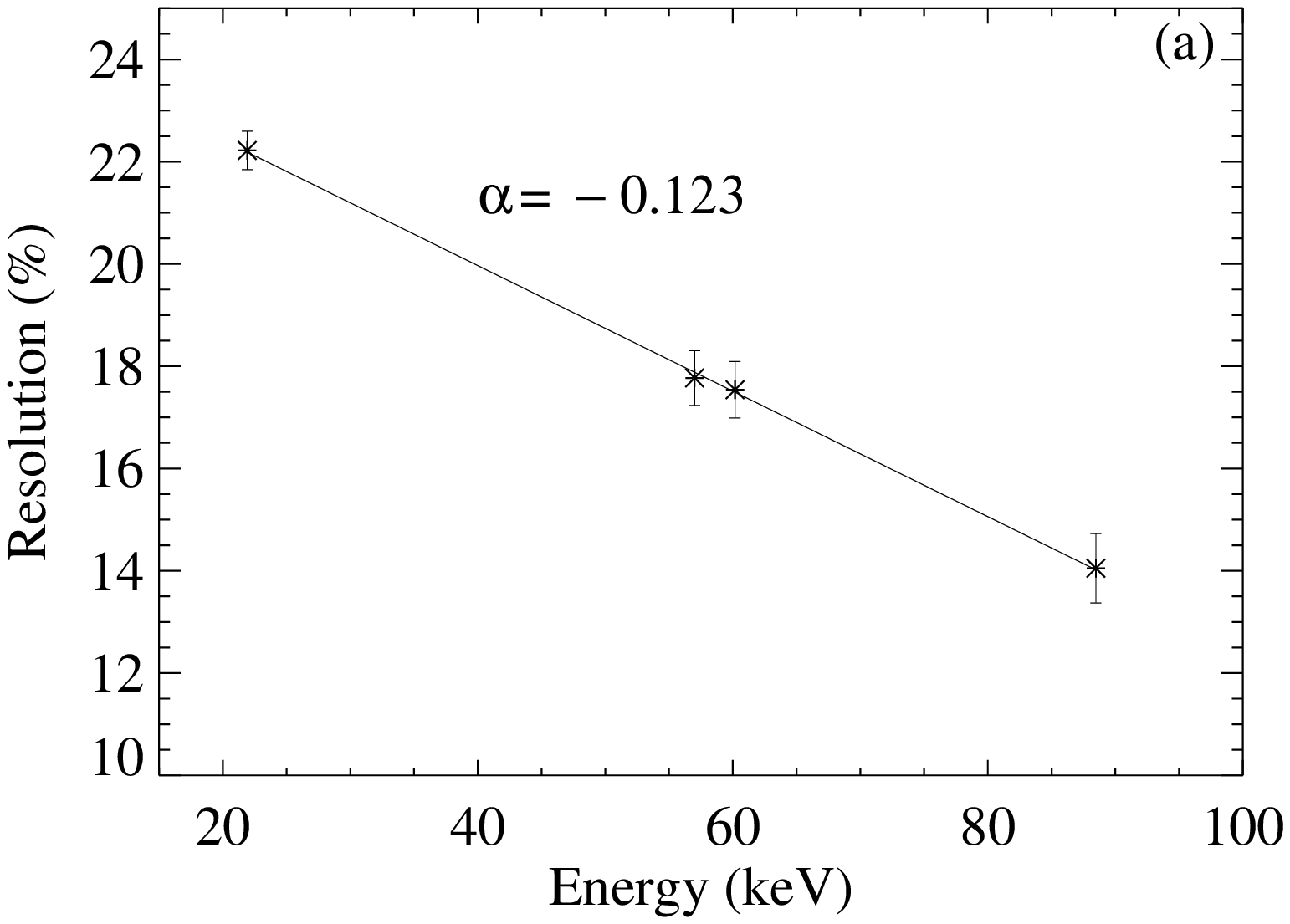}
\hskip 0.01cm
\includegraphics[width=5.2cm,height=4.8cm,angle=0]{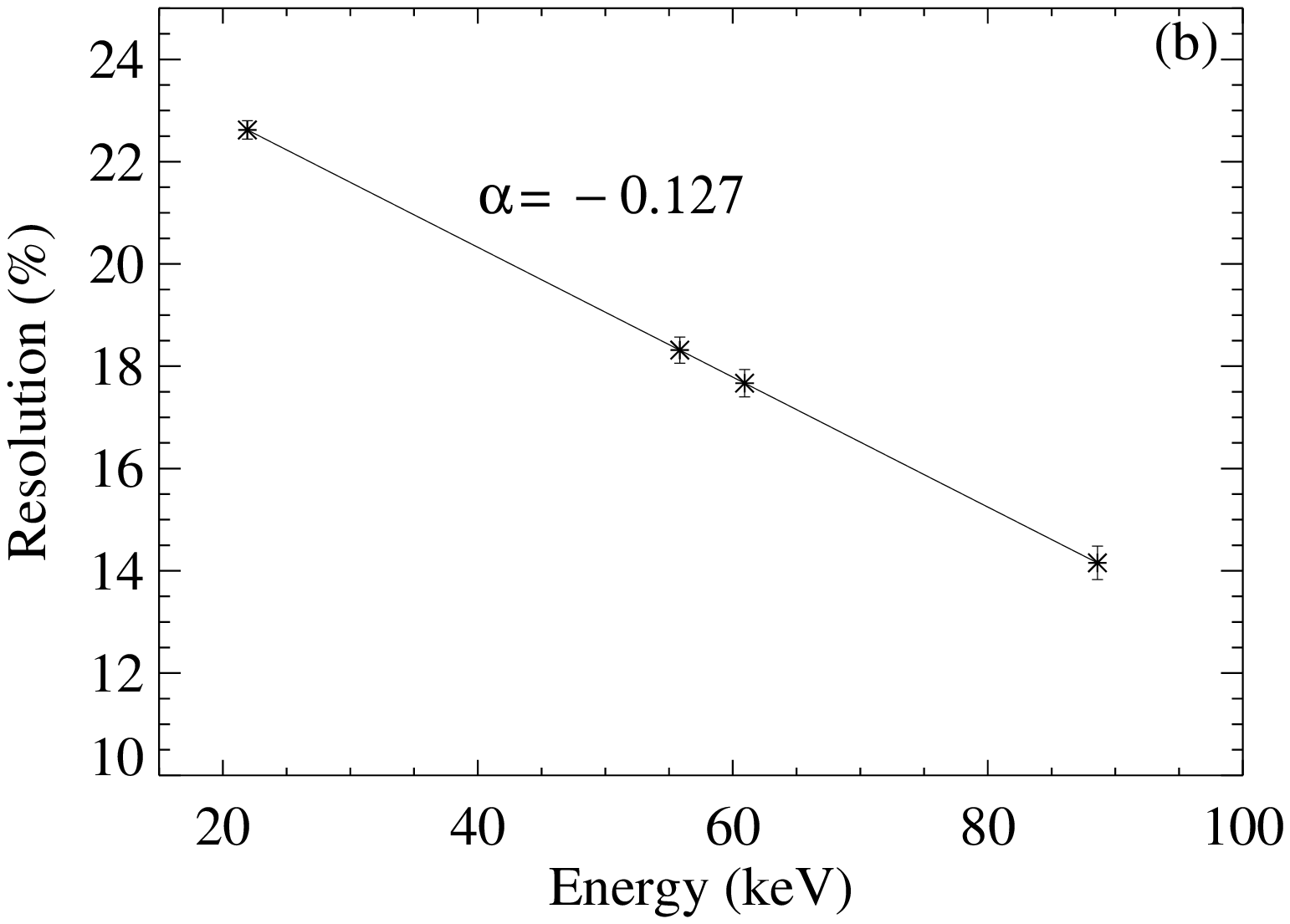}}}
\hskip 0.01cm
\caption{The variation of energy resolution of NaI spectrum as a function of energy for both
instruments (left panel for RT-2/S and right panel for RT-2/G).}
\label{kn : fig10(ab)}
\end {figure}

In the next section, we discuss the G1-CsI and G2 energy calibration procedure.

\subsection{Laboratory Calibration of RT-2/S \& RT-2/G Spectra}


As already discussed, both instruments have three separate energy spectra: G1-NaI, G1-CsI 
and G2 with different characteristics. The G1-NaI and G1-CsI spectra cover 1024 channels, whereas 
G2 spectrum covers only 256 channels. PSD technique is used to discriminate the events that are
detected in NaI and CsI crystals for G1 amplifier only. The PSD spectrum of both instruments are
shown in Figures 11(a) and 12(a) (corresponding PS cut values are marked in the respective 
Figures). A high gain amplifier G2 is also used to register the high energy events in CsI 
crystal to have G2 spectrum.  

\begin {figure}[h]
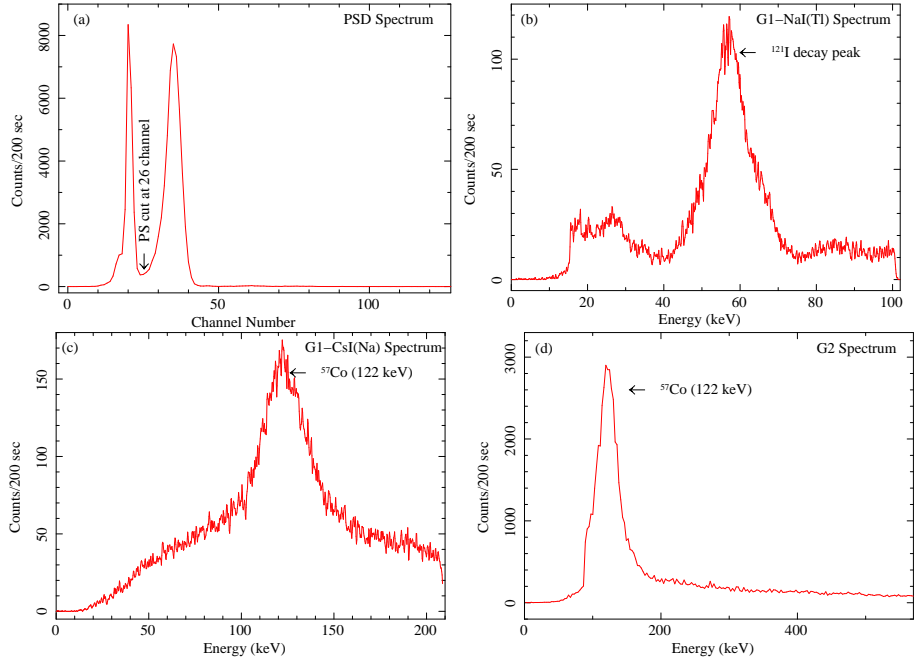

\vbox{
\vskip 0.0 cm
\centering{
\includegraphics[width=4.3cm,angle=270]{fig11a.ps}
\hskip 0.01cm
\includegraphics[width=4.3cm,angle=270]{fig11b.ps}}
\vskip 0.02cm
\centering{
\includegraphics[width=4.3cm,angle=270]{fig11c.ps}
\hskip 0.01cm
\includegraphics[width=4.3cm,angle=270]{fig11d.ps}}}
\vskip 0.06cm
\caption{Four spectra of RT-2/S payload: (a) PSD spectrum, (b) G1-NaI(Tl) spectrum, (c) G1-CsI(Na) 
spectrum and (d) G2 spectrum. Calibration source peak of $^{57}$Co (122 keV) is detected at G1-CsI 
and G2 spectra.}
\label{kn : fig10(a-d)}
\end{figure}



In order to calibrate the G1-CsI and G2 energy scale, we used a 122 keV $^{57}$Co
line. During the laboratory tests, the G1-CsI and G2 gains were changed by
varying the HV, LLD and ULD voltages in order to: i) identify the PH
values of the 122 keV line in channel number and ii) study the G1-CsI and G2 gain
stability.

The measured G1-CsI and G2 gain values are 0.2289 keV/Chan and 2.732 keV/Chan
for RT-2/S and 0.2279 keV/Chan and 3.948 keV/Chan for RT-2/G, respectively.
The G1-CsI and G2 energy converted spectra are used to measure the energy
resolution of the 122 keV emission line for both instruments. The measured
G1-CsI and G2 values, which are consistent with previous experiments (such as
BeppoSAX, and RXTE using similar detectors), are 19.79\% and 25.73\% for RT-2/S
and 20.69\% and 18.65\% for RT-2/G, respectively.

Apart from the energy scale calibration, it is also essential to have
the G1-CsI and G2 background for both instruments. The
in-flight background calibration is carried out using a $^{57}$Co radio-active
source (122 keV) placed in one of the slats of the collimator.

The energy calibrated background spectra of G1-NaI, G1-CsI and G2 for both instruments are shown 
in Figure 11(b,c,d) and in Figure 12(b,c,d). The G1-NaI spectrum shows the signature of 
$^{121}$I decay peak (due to $^{57}$Co, discussed in the previous section) at 57.19 keV and at 
56.05 keV for RT-2/S and RT-2/G respectively. The background spectra G1-CsI and G2 are dominated 
by the line emission of 122 keV of $^{57}$Co over the background continuum. The large `continuum' 
in G1-CsI spectrum is due to high energy background photons that are detected in CsI crystal 
in wider energy range of $\sim$30 keV to 210 keV along with the broad emission feature of 122 
keV line (energy resolution 19.79\% for RT-2/S and 20.69\% for RT-2/G). 
The G2 spectra of both instruments are from extremely high energy photons (as detected in the 
laboratory). The 122 keV line intensity is strong enough as compared to the high energy background photons, which
causes the G2 spectra a `tail' like continuum. 

\begin {figure}[h]
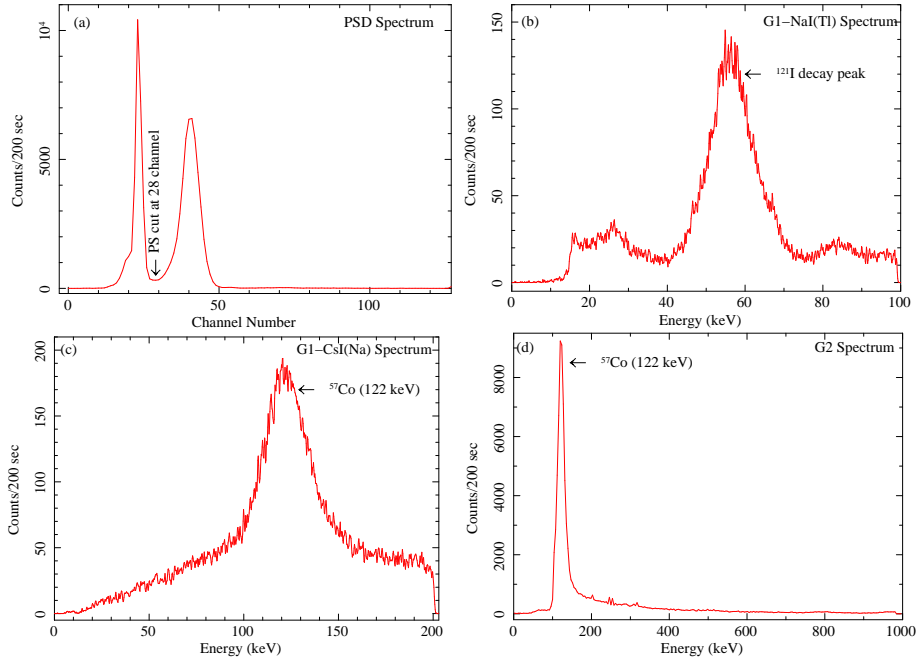

\vbox{
\vskip 0.0 cm
\centering{
\includegraphics[width=4.3cm,angle=270]{fig12a.ps}
\hskip 0.01cm
\includegraphics[width=4.3cm,angle=270]{fig12b.ps}}
\vskip 0.02cm
\centering{
\includegraphics[width=4.3cm,angle=270]{fig12c.ps}
\hskip 0.01cm
\includegraphics[width=4.3cm,angle=270]{fig12d.ps}}}
\vskip 0.06cm
\caption{Four spectra of RT-2/G payload: (a) PSD spectrum, (b) G1-NaI(Tl) spectrum, (c) G1-CsI(Na) 
spectrum and (d) G2 spectrum. Calibration source peak of $^{57}$Co (122 keV) is detected at G1-CsI 
and G2 spectra.}
\label{kn : fig11(a-d)}
\end{figure}

Although the `phoswich' technique (discussed in \S 3) is useful
for background rejection for charge particles, but we could not perform any specific test in 
the ground (laboratory test) with charged particles in order to test the particle background 
rejection process. 

\section{Concluding remarks}

In this paper, we described the complete functionality (e.g., HV calibration, Thermistor 
calibration, stability of PS and PH variation, validity of spectral data, health parameters 
etc.) during various environmental tests of RT-2 payloads (RT-2/S \& RT-2/G) of RT-2 Experiment 
onboard the CORONAS-PHOTON mission which has been launched on January 30, 2009. Laboratory results, 
which include the space environmental tests (e.g., vibration, thermal, vacuum, emi, src tests etc.) 
and the functionality tests 
suggest that both instruments are qualified for space use. All these tests were carried out
in a period of three months. We found that the scientific results based on functionality
test are stable for the entire period of the testing phase.
In fact, both instruments (RT-2/S and RT-2/G) are identical (except their FOV and in low energy 
working range) as far as the configuration details are concerned. However, the
basic differences in their spectral characteristics 
(e.g., amplitude, shape, energy ranges etc.) are due to the 
difference in technical settings of the operational conditions (e.g., HV, LLD/ULD variation, PSD
setting etc.). The energy ranges (channel boundary) of the 
amplifiers (G1 and G2) of both instruments are 
very wide and would be quite useful. During the onboard operation, different 
channel boundaries (energy ranges) would be set
to understand the background rejection effect for both instruments (payloads).


To study the particle background rejection process, a `robust' logic 
was implemented inside the FPGA to process any `event' that is detected 
in the crystals. PSD technique was used to separate out the event registered in NaI or CsI 
crystal. The photons with higher energy ($\ge$ 100 keV) are mostly detected in CsI crystal. Apart 
from the PSD technique, the in-built logic inside the FPGA also can separate out the
high energy photons ($\ge$ 40 keV) which are detected in CsI crystal of G2 amplifier (G2 spectrum).
                                                                                


During the commissioning phase of both instruments, we found that the initial results (Nandi et 
al. 2009) are in agreement with the results obtained in the ground condition. 
In its lifetime ($\sim$ 3-5 years), RT-2 will cover the peak of the $24^{th}$ solar cycle. So 
far, the solar activity has not really taken off. Only a few weak flares which occurred have been 
successfully detected by RT-2/S and RT-2/G, and the results will be reported elsewhere.

\noindent {\bf Acknowledgments:}
DD thanks CSIR/NET scholarship and TBK thanks RT-2/SRF fellowship which supported their research 
work. The authors are thankful to scientists, engineers and technical staffs from 
TIFR/ICSP/VSSC/ISRO-HQ for various supports during RT-2 related experiments. The authors also
thank the anonymous referee for the very detailed comments to improve the quality of the manuscript.

{}

\end{document}